\documentclass[sigconf]{acmart}
\usepackage{subfigure}

\newcommand{\be}{\begin{equation}}
\newcommand{\ee}{\end{equation}}
\newcommand{\bea}{\begin{eqnarray}}
\newcommand{\eea}{\end{eqnarray}}
\newcommand{\ba}{\begin{array}}
\newcommand{\ea}{\end{array}}
\newcommand{\bsm}{\begin{small}}
\newcommand{\esm}{\end{small}}
\newcommand{\nid}{\noindent}
\newcommand{\bfa}{\mathbf{a}}

\newcommand{\maximize}{\textrm{maximize}}

\newcommand{\at}{\varphi^{(\mathrm{tx})}_{i}}
\newcommand{\et}{\vartheta^{(\mathrm{tx})}_{i}}
\newcommand{\ar}{\varphi^{(\mathrm{rx})}_{j}}
\newcommand{\er}{\vartheta^{(\mathrm{rx})}_{j}}

\AtBeginDocument{%
  }

\copyrightyear{2024}
\acmYear{2024}
\setcopyright{acmlicensed}\acmConference[ACM MobiCom '24]{The 30th Annual International Conference on Mobile Computing and Networking}{November 18--22, 2024}{Washington D.C., DC, USA}
\acmBooktitle{The 30th Annual International Conference on Mobile Computing and Networking (ACM MobiCom '24), November 18--22, 2024, Washington D.C., DC, USA}
\acmDOI{10.1145/3636534.3698223}
\acmISBN{979-8-4007-0489-5/24/11}

\acmConference[ISACom '24]{3rd ACM MobiCom Workshop on Integrated Sensing and Communications Systems}{November 18, 2024}{Washington, DC}

\begin{document}

\title{Heterogeneous Graph Neural Network for Cooperative ISAC Beamforming in Cell-Free MIMO Systems}

\author{Zihuan Wang}
\email{zihuanwang@ece.ubc.ca}
\affiliation{%
  \institution{The University of British Columbia}
  \city{Vancouver}
  \country{Canada}
}

\author{Vincent W.S. Wong}
\email{vincentw@ece.ubc.ca}
\affiliation{%
  \institution{The University of British Columbia}
  \city{Vancouver}
  \country{Canada}
}


\begin{abstract}
Integrated sensing and communication (ISAC) is one of the usage scenarios for the sixth generation (6G) wireless networks. 
In this paper, we study cooperative ISAC in cell-free multiple-input multiple-output (MIMO) systems, where multiple MIMO access points (APs) collaboratively provide communication services and perform multi-static sensing. 
We formulate an optimization problem for the ISAC beamforming design, which maximizes the achievable sum-rate while guaranteeing the sensing signal-to-noise ratio (SNR) requirement and total power constraint. Learning-based techniques are regarded as a promising approach for addressing such a nonconvex optimization problem.  
By taking the topology of cell-free MIMO systems into consideration, we propose a heterogeneous graph neural network (GNN), namely SACGNN, for ISAC beamforming design. The proposed SACGNN framework models the cell-free MIMO system for cooperative ISAC as a heterogeneous graph and employs a transformer-based heterogeneous message passing scheme to capture the important information of sensing and communication channels and propagate the information through the graph network. Simulation results demonstrate the performance gain of the proposed SACGNN framework over a conventional null-space projection based scheme and a deep neural network (DNN)-based baseline scheme. 
\end{abstract}




\maketitle

\section{Introduction}

Integrated sensing and communication (ISAC) is envisioned to play an indispensable role in the sixth generation (6G) wireless networks, supporting immersive communications and pervasive sensing, and facilitating paradigm shifts towards the metaverse \cite{Nuria, FL-JSAC22, metaverse,Tataria}. Benefiting from the terahertz frequency bands and large antenna arrays, ISAC systems can provide sensing and communication functionalities simultaneously through multiple-input multiple-output (MIMO) beamforming, where the sensing and communication modules share the same infrastructure and spectrum resources. Thus, ISAC can reduce hardware cost and improve the spectral and energy efficiencies for wireless communication networks. 

Extensive efforts have been devoted to theoretical analysis and testbed development for ISAC beamforming design \cite{FL-JSAC22,CD-mobicom,TX-mobicom,ICC,JSTSP}. These works focus on single base station scenarios. In order to further enhance the spatial degrees of freedom (DoFs), cooperative communication and multi-static sensing are crucial. This motivates the investigation of cooperative ISAC in cell-free MIMO systems. In such cell-free MIMO systems, multiple access points (APs) collaboratively provide seamless communication service and receive uncorrelated sensing observations. The distributed APs are connected to a central processing unit (CPU), which facilitates joint processing between the APs \cite{ZB-TWC,YH-TVT22,UD-Asilomar,SL-WCNC}.
In \cite{ZB-TWC,YH-TVT22}, power allocation is investigated to maximize the sensing performance
while ensuring that the communication performance meets its requirements. 
Cooperative ISAC beamforming design in cell-free MIMO networks is investigated in \cite{UD-Asilomar}. A semidefinite relaxation
(SDR) based algorithm is proposed for transmit beamforming design which aims to maximize the sensing signal-to-noise ratio
(SNR) while satisfying the communication  signal-to-interference-plus-noise ratio
(SINR) requirement. 
In \cite{SL-WCNC}, the mode selection of APs is studied for joint communication and multi-static sensing, where each AP can operate either as a transmitter or receiver. 

The aforementioned works rely on alternating optimization algorithms for power allocation and beamforming design, which can incur a high computational complexity. Recently, deep learning techniques, which train the deep neural networks (DNNs) offline and then deploy the trained models for online optimization, have proven to reduce the computational complexity of solving optimization problems in wireless networks.
Among the various types of DNNs, graph neural networks (GNNs) \cite{HGT,HetGNN,YS-TWC23} have demonstrated significant potential for modeling structured data. This makes them particularly well-suited for solving optimization problems in wireless networks with specific architectures, such as cell-free systems. 
Existing GNNs for wireless networks typically account only for communication functionality, and they primarily focus on extracting and passing the message carried by communication channels. ISAC systems, on the other hand, also involve sensing information, which should be incorporated into the graph modeling and message passing procedures.

In this paper, we investigate cooperative ISAC beamforming in cell-free MIMO systems. We develop a heterogeneous GNN, which is called sensing and communication GNN (SACGNN), for ISAC beamforming design to maximize the achievable sum-rate while guaranteeing the sensing SNR and total power constraints. The proposed SACGNN framework models the cell-free MIMO system as a heterogeneous graph, with AP antennas and users being different types of nodes. Then, we apply a transformer-based message passing scheme, which extracts important information from sensing and communication channels with attention mechanisms. 
We conduct simulations to evaluate the performance of our proposed SACGNN framework. The results demonstrate the effectiveness of the proposed approach over a conventional null-space projection based scheme and a DNN-based baseline scheme.

\section{System Model \label{sec:system model}}

\begin{figure}[t]
\centering
\includegraphics[width=2.4 in]{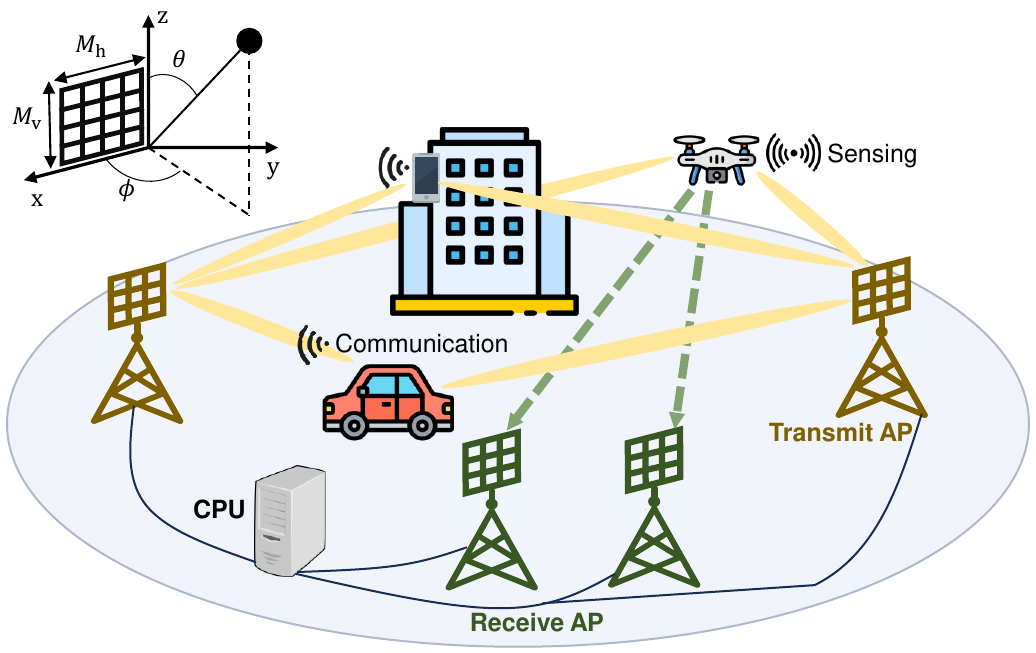}

\caption{The system model of cooperative ISAC transmission in a cell-free MIMO system. The transmit APs jointly communicate with multiple users and sense a target through beamforming. The receive APs obtain the echo signal reflected by the target. The transmit and receive APs are connected to a CPU through fronthaul links.}
\Description{System model.}
\label{fig:system model}\vspace{-0.4 cm}
\end{figure}

We consider a cell-free MIMO system for joint communication and multi-static sensing. There are $N_{\mathrm{T}}$ transmit APs and $N_{\mathrm{R}}$ receive APs, which are synchronized and connected to a CPU via fronthaul links, as shown in Fig. \ref{fig:system model}. Each AP is equipped with a uniform planar array (UPA) with $M = M_{\mathrm{v}}M_{\mathrm{h}}$ antennas, where $M_{\mathrm{v}}$ and $M_{\mathrm{h}}$ denote the number of antennas in the vertical and horizontal dimensions, respectively. For a half-wavelength-spaced UPA, the vertical and horizontal beam steering vectors are respectively given by
\bea
&&\hspace{-1 cm}\mathbf{b}_{\mathrm{v}}(\theta) = \frac{1}{\sqrt{M_{\mathrm{v}}}} \left[1~ e^{-j\pi\cos\theta}~ \cdots ~ e^{-j(M_{\mathrm{v}}-1)\pi\cos\theta} \right]^T, \\
&&\hspace{-1 cm}\mathbf{b}_{\mathrm{h}}(\phi, \theta) = \frac{1}{\sqrt{M_{\mathrm{h}}}} \left[1~ e^{-j\pi\sin\theta\cos\phi}~ \cdots ~ e^{-j(M_{\mathrm{h}}-1)\pi\sin\theta\cos\phi} \right]^T, 
\eea 
where $\phi$ and $\theta$  are the azimuth and elevation angles, respectively. The beam steering vector can be expressed as $\mathbf{a}(\phi, \theta) \in \mathbb{C}^{M}$:
\bea
\mathbf{a}(\phi, \theta) = \mathbf{b}_{\mathrm{h}}(\phi, \theta) \otimes \mathbf{b}_{\mathrm{v}}(\theta).
\eea 
The $N_{\mathrm{T}}$ transmit APs jointly communicate with  $K$ users through beamforming and steer another beamformer toward a specific location for target detection. Note that the beamformers for communication and sensing share the same waveform and time-frequency resources. On the other hand, the $N_{\mathrm{R}}$ receive APs obtain the reflected echo signals, which are used to determine whether there is a target at that particular location or not. 


\subsection{Signal Model}

We consider ISAC downlink transmission, where the transmit APs collaboratively send $K$ communication data streams  $\{s_k\}_{k=1}^K$ to $K$ users and transmit another data stream $s_0$ for target detection. The total number of streams is equal to $K + 1$. 
Each transmitted symbol has unit power, i.e., $\mathbb{E}\{|s_k|^2\} = 1$, for $k=0, \ldots, K$. We define $\mathbf{s} = [s_0 ~ \cdots ~ s_K]^T \in \mathbb{C}^{K+1}$ as the transmitted streams. Similar to \cite{UD-Asilomar}, we assume the streams are statistically independent, i.e., $\mathbb{E}\{\mathbf{s}\mathbf{s}^H\} = \mathbf{I}_{K+1}$. 
Let $\mathbf{x}_i \in \mathbb{C}^{M}$ denote the transmitted signal at the $i$-th transmit AP.  It can be expressed as  
\bea
\mathbf{x}_i = \underbrace{\sum_{k=1}^K \mathbf{f}_{i,k}s_k}_\text{Communication} + \underbrace{ \mathbf{f}_{i,0}s_0}_\text{Sensing} = \sum_{k=0}^K \mathbf{f}_{i,k}s_k = \mathbf{F}_i\mathbf{s}, 
\eea 
where $\mathbf{f}_{i,k}\in \mathbb{C}^{M}$ is the beamforming vector of the $i$-th transmit AP for the $k$-th data stream. We define $\mathbf{F}_i = [\mathbf{f}_{i,0} ~ \cdots ~ \mathbf{f}_{i,K}] \in \mathbb{C}^{M\times (K+1)}$ as the beamforming matrix at the $i$-th transmit AP. 
The transmit power at the $i$-th transmit AP  is given by 
$P_i =  \|\mathbf{F}_{i}\|^2_F, ~~ i = 1, \ldots, N_{\mathrm{T}}$. 
Each transmit AP has a constraint on the total power consumption, i.e., $P_i \leq P_{\max}$, for $i = 1, \ldots, N_{\mathrm{T}}$, where $P_{\max}$ is the maximum transmit power. 

\subsection{Communication Model}
\allowdisplaybreaks

Let $\mathbf{h}_{i,k} \in \mathbb{C}^{M}$ denote the channel vector between the $i$-th transmit AP and the $k$-th user, $i = 1, \ldots, N_{\mathrm{T}}$, $k = 1, \ldots, K$. Assuming there is a line-of-sight (LoS) link between each transmit AP and user, the channel vector $\mathbf{h}_{i,k}$ can be expressed as 
\bea
&&\hspace{-0.7 cm} \mathbf{h}_{i,k} = \beta_{i,k}\mathbf{a}(\phi_{i,k},\theta_{i,k}), ~ i = 1, \ldots, N_{\mathrm{T}}, ~k = 1, \ldots, K, 
\eea 
where $\beta_{i,k}$ denotes the complex channel gain, and it follows a Gaussian distribution with zero mean and variance $\zeta_{i,k}^2$. $\phi_{i,k}$ and $\theta_{i,k}$ are the angles of departure (AoDs) in the azimuth and elevation domains of the $k$-th user observed at the $i$-th transmit AP, respectively. 
By stacking the channel vectors between the $k$-th user and all the transmit APs, we define $\mathbf{h}_k  = [\mathbf{h}^T_{1,k}~ \cdots ~ \mathbf{h}^T_{N_{\mathrm{T}},k}]^T \in \mathbb{C}^{MN_{\mathrm{T}}}$.
By stacking the beamforming vectors of all the transmit APs for the $k$-th user, we define
$
\mathbf{f}_k = [\mathbf{f}^T_{1,k}~ \ldots ~ \mathbf{f}^T_{N_{\mathrm{T}},k}]^T \in \mathbb{C}^{MN_{\mathrm{T}}}$.
Then, the received downlink signal at the $k$-th user can be expressed as follows:
\bea
&&\hspace{-0.9 cm} y^{(\mathrm{c})}_k =  \sum_{i=1}^{N_{\mathrm{T}}} \mathbf{h}^H_{i,k} \mathbf{x}_i + n_k \\ &&\hspace{-0.65 cm} = \underbrace{\mathbf{h}_k^H\mathbf{f}_ks_k}_\text{Desired signal} + \hspace{-0.1 cm}\underbrace{\sum_{m=1, m\neq k}^K \mathbf{h}_k^H\mathbf{f}_{m}s_m}_\text{Multi-user interference} + \hspace{-0.1 cm}\underbrace{\mathbf{h}_k^H\mathbf{f}_{0}s_0}_\text{Sensing interference} + \hspace{-0.1 cm}\underbrace{n_k}_\text{Noise},
\eea 
where $n_k$ denotes white Gaussian noise of the $k$-th user with zero mean and variance $\sigma_k^2$. Then, the SINR of the $k$-th user, i.e., $\gamma^{(\mathrm{c})}_k$, can be obtained as follows:
\bea
\gamma^{(\mathrm{c})}_k = \frac{|\mathbf{h}_k^H\mathbf{f}_k|^2}{\sum_{m=1, m\neq k}^K |\mathbf{h}_k^H\mathbf{f}_m|^2 + |\mathbf{h}^H_k\mathbf{f}_0|^2 + \sigma_k^2}.
\eea 
The achievable sum-rate of all $K$ users is given by
\bea
R = \sum_{k=1}^K \log_2\left(1+\gamma_k^{(\mathrm{c})}\right). 
\eea 

\subsection{Sensing Model}
If a target exists at the point of interest, then the transmitted downlink signals will be reflected by the target and the reflected echo signal will be collected by the receive APs. Assuming a LoS path exists between each AP and the target, the echo signal obtained at the $j$-th receive AP, where $j = 1, \ldots, N_{\mathrm{R}}$,  is given by\footnote{There are also direct communication links between the transmit and receive APs. We assume that these direct links are known prior to sensing, and their effects can be removed from the received echo signal at each receive AP \cite{ZB-TWC}. Moreover, there may also exist multipath components. Similar to \cite{ZB-TWC,UD-Asilomar}, we assume the contribution of the multipath components is small and can be ignored for simplicity.}
\bea
\mathbf{y}^{(\mathrm{s})}_j = \sum_{i=1}^{N_{\mathrm{T}}} \lambda_{i,j}\bfa(\ar, \er)\bfa^H(\at, \et)\mathbf{x}_i  + \mathbf{z}_j,  \label{eq:echo1}
\eea 
where $\lambda_{i,j}$ is an unknown complex sensing channel gain. $\lambda_{i,j}$ includes the effects due to path loss and radar cross section of the target. It follows the Gaussian distribution with zero mean and variance $\chi^2_{i,j}$. $\at$ and  $\et$ correspond to the azimuth and elevation AoDs of the target observed at the $i$-th transmit AP, respectively. $\ar$ and  $\er$ correspond to the azimuth and elevation angles of arrival (AoAs) of the target observed at the $j$-th receive AP, respectively.  
$\mathbf{z}_j\in \mathbb{C}^{M}$ is the noise at the $j$-th receive AP, which follows the complex Gaussian distribution with zero mean and variance of $\xi_{j}^2\mathbf{I}_M$. 

We define matrix $\mathbf{A}_{i,j} \overset{\Delta}{=} \lambda_{i,j}\bfa(\ar, \er)\bfa^H(\at, \et) \in \mathbb{C}^{M\times M}$.  
A compact form of the obtained echo signal (\ref{eq:echo1}) at the $j$-th receive AP can be expressed as 
\bea
\mathbf{y}_j^{(\mathrm{s})} =  \mathbf{A}_j\mathbf{x} + \mathbf{z}_j, \label{eq:echo}
\eea 
where matrix $\mathbf{A}_j = [\mathbf{A}_{1,j}~ \cdots ~\mathbf{A}_{N_{\mathrm{T}},j}] \in \mathbb{C}^{M \times N_{\mathrm{T}}M}$ and vector $\mathbf{x} = [\mathbf{x}^T_1~ \cdots~ \mathbf{x}^T_{N_{\mathrm{T}}}]^T \in \mathbb{C}^{N_{\mathrm{T}}M}$. Then, each receive AP forwards the echo signal (\ref{eq:echo}) to the CPU. The concatenated echo signal obtained from all the receive APs is given by $\mathbf{y}^{(\mathrm{s})} = [(\mathbf{y}^{(\mathrm{s})}_1)^T~ \cdots ~ (\mathbf{y}^{(\mathrm{s})}_{N_{\mathrm{R}}})^T]^T \in \mathbb{C}^{N_{\mathrm{R}}M}$. It can be expressed as 
\bea 
\mathbf{y}^{(\mathrm{s})} =  \mathbf{A}\mathbf{x} + \mathbf{z},
\label{eq:compact y}
\eea 
where $\mathbf{A} = [\mathbf{A}^T_1~\cdots~ \mathbf{A}^T_{N_{\mathrm{R}}}]^T \in \mathbb{C}^{N_{\mathrm{R}}M \times N_{\mathrm{T}}M}$ and $\mathbf{z} = [\mathbf{z}^T_1~ \cdots~ \mathbf{z}^T_{N_{\mathrm{R}}}]^T \in \mathbb{C}^{N_{\mathrm{R}}M}$. 
The sensing SNR for target detection is given by
\bea 
\gamma^{(\mathrm{s})} = \frac{\mathbb{E}\{\|\mathbf{A}\mathbf{x}\|^2\}}{\mathbb{E}\{\|\mathbf{z}\|^2\}}  
= \frac{\sum_{j=1}^{N_{\mathrm{R}}}\sum_{i=1}^{N_{\mathrm{T}}}\chi^2_{i,j}\big\|\widetilde{\mathbf{A}}_{i,j} \mathbf{F}_i\big\|^2_F}{\sum_{j=1}^{N_{\mathrm{R}}}\xi_{j}^2}, 
\eea 
where we define $\widetilde{\mathbf{A}}_{i,j} \overset{\Delta}{=}  \bfa(\ar, \er)\bfa^H(\at, \et) \in \mathbb{C}^{M\times M}$.

\subsection{Problem Formulation}

In this paper, we consider cooperative ISAC beamforming design to maximize the achievable sum-rate while guaranteeing the sensing SNR requirement and the total power constraint at each transmit AP. The optimization problem can be formulated as follows:
\begin{subequations}
\begin{align}
& \hspace{-0.1 cm}\underset{\{\mathbf{F}_{i}\}_{i=1}^{N_{\mathrm{T}}}}{\maximize}~ R
\\
&\hspace{0.0 cm} \textrm{subject to} ~~
\gamma^{(\mathrm{s})} \geq \gamma_{\min}, \label{eq:sense constraint}
\\ 
&\hspace{1.3 cm} \|\mathbf{F}_{i}\|^2_F \leq P_{\max},~~  i=1, \ldots, N_{\mathrm{T}},\label{eq:power constraint}
\end{align}\label{eq:original problem}
\end{subequations}

\vspace{-0.3 cm}

\nid where $\gamma_{\min}$ is the minimum SNR requirement for target sensing. Note that the optimal solution to problem (\ref{eq:original problem}) is intractable due to the nonconvexity of the objective function. 
Motivated by the recent success of applying learning-based techniques to solve nonconvex optimization problems, we propose a SACGNN framework in the next section, which leverages graph learning to effectively capture the structural information of cell-free MIMO systems and determine the beamformers $\{\mathbf{F}_{i}\}_{i=1}^{N_{\mathrm{T}}}$. 

\section{Proposed SACGNN Framework \label{sec:proposed}}

In this section, we introduce the SACGNN framework for cooperative ISAC beamforming design. The proposed SACGNN framework models the cell-free MIMO system as a heterogeneous graph and uses a transformer-based heterogeneous message passing scheme to extract important sensing and communication features.

\subsection{Heterogeneous ISAC Graph Modeling}

\begin{figure}[t]
\centering
\includegraphics[width=2.3 in]{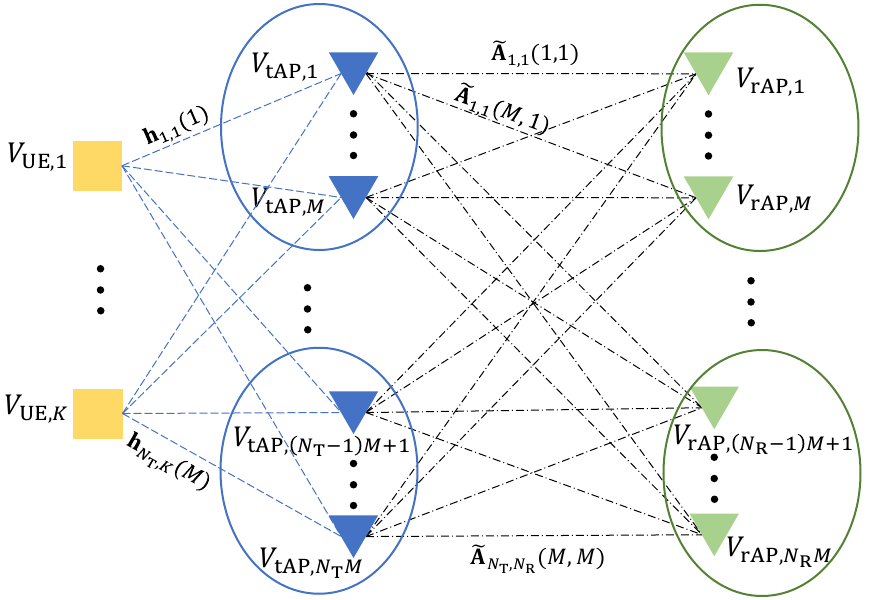}
\vspace{0mm}
\caption{The heterogeneous graph model of a cell-free MIMO system for cooperative ISAC.}
\Description{SACGNN.}
\label{fig:graph model}
\vspace{-0.2 cm}
\end{figure}

The considered cell-free MIMO system for cooperative ISAC can be modeled as a heterogeneous graph, denoted as $\mathcal{G} = (\mathcal{V}, \mathcal{E})$, which consists of a set of nodes $\mathcal{V}$ and the corresponding set of edges $\mathcal{E}$. 
The graph model consists of three types of nodes. The first type is the   
transmit AP antenna (tAP) which sends ISAC signals. The second type is the receive AP antenna (rAP) which obtains the reflected echo signals for sensing. The third type is the user equipment (UE) which receives communication service. We define the set of node types as $\mathcal{A} = \{\mathrm{tAP}, \mathrm{rAP}, \mathrm{UE}\}$, and function $\psi(V): \mathcal{V}\rightarrow \mathcal{A}$ which maps a node $V\in\mathcal{V}$ to its node type in $\mathcal{A}$. 
Accordingly, the set of nodes can be expressed as $\mathcal{V} = \mathcal{V}_{\mathrm{tAP}}\cup \mathcal{V}_{\mathrm{rAP}}\cup \mathcal{V}_{\mathrm{UE}}$, which consists of tAP node set $\mathcal{V}_{\mathrm{tAP}} = \{V_{\mathrm{tAP,1}}, \ldots, V_{\mathrm{tAP}, N_{\mathrm{T}}M}\}$, rAP node set $\mathcal{V}_{\mathrm{rAP}} = \{V_{\mathrm{rAP,1}}, \ldots, V_{\mathrm{rAP},N_{\mathrm{R}}M}\}$, and UE node set $\mathcal{V}_{\mathrm{UE}} =  \{V_{\mathrm{UE,1}}, \ldots, V_{\mathrm{UE},K}\}$. 
Each pair of tAP and UE nodes is connected by an undirected edge, with the channel coefficient being their edge feature. In particular, as shown in Fig. \ref{fig:graph model}, $\mathbf{h}_{i,k}(m)$ serves as the edge feature between node $V_{\mathrm{tAP}, (i-1)M+m}$ and node $V_{\mathrm{UE}, k}$. 
Note that $\mathbf{h}_{i,k}(m)$ represents the $m$-th element of vector $\mathbf{h}_{i,k}$, where $i=1, \ldots, N_{\mathrm{T}}, k = 1, \ldots, K, m=1, \ldots, M$.  
The tAP nodes are also connected with the rAP nodes. $\widetilde{\mathbf{A}}_{i,j}(m,n)$ serves as the edge feature between node $V_{\mathrm{tAP}, (i-1)M+n}$ and node $V_{\mathrm{rAP}, (j-1)M+m}$. Note that $\widetilde{\mathbf{A}}_{i,j}(m,n)$ denotes the  element in the $m$-th row and $n$-th column of matrix $\widetilde{\mathbf{A}}_{i,j}$, for 
$i=1, \ldots, N_{\mathrm{T}}, j = 1, \ldots, N_{\mathrm{R}}, m=1, \ldots, M, n = 1, \ldots, M$.

\subsection{Heterogeneous
Message Passing}

Consider a SACGNN framework with $L$ hidden layers and denote  $\mathbf{g}^{(l)}[V]$ as the hidden state of the $l$-th layer of node $V \in \mathcal{V}$. Denote $\mathcal{N}(V)$ as the set of neighboring nodes that have edges connected to node $V \in \mathcal{V}$. 
In each layer, each node updates its own hidden state based on the aggregated message passed from its neighboring nodes. Due to node heterogeneity, the features related to various types of nodes may fall into different feature spaces. 
It is crucial to handle the diverse feature information and capture the important features. In this work, we apply a transformer \cite{transformer} based scheme to determine the attention scores for the set of neighbors, which takes the importance of node heterogeneity into consideration. 

In particular, we first initialize the hidden state for each node based on the edge features shown in Fig. \ref{fig:graph model}.  For a tAP node $V_{\mathrm{tAP}, i} \in \mathcal{V}_{\mathrm{tAP}}$, where $i = 1, \ldots, N_{\mathrm{T}}M$, the initial hidden state $\mathbf{g}^{(1)}[V_{\mathrm{tAP}, i}] \in \mathbb{C}^{K + N_{\mathrm{R}}M}$ for the first layer combines the edge features associated with the $i$-th tAP node. It is given by
\bea
&&\hspace{-0.6 cm}\mathbf{g}^{(1)}[V_{\mathrm{tAP}, i}] = \Big[\mathbf{h}_{\lfloor \frac{i}{M} \rfloor, 1}(i - M\lfloor (i-1)/M \rfloor), \cdots, \nonumber \\
&& \hspace{-0.0 cm} \mathbf{h}_{\lfloor \frac{i}{M} \rfloor, K}(i - M\lfloor (i-1)/M \rfloor), \widetilde{\mathbf{A}}_{\lfloor \frac{i}{M} \rfloor, 1} (i - M\lfloor (i-1)/M \rfloor, 1), \nonumber \\
&& \hspace{-0.6 cm} \cdots, \widetilde{\mathbf{A}}_{\lfloor \frac{i}{M} \rfloor, 1} (i - M\lfloor (i-1)/M \rfloor, M), \widetilde{\mathbf{A}}_{\lfloor \frac{i}{M} \rfloor, 2} (i - M\lfloor (i-1)/M \rfloor, 1),\nonumber \\
&& \hspace{-0.6 cm}  \cdots,  \widetilde{\mathbf{A}}_{\lfloor \frac{i}{M} \rfloor, N_{\mathrm{R}}} (i - M\lfloor (i-1)/M \rfloor, M)\Big]^T,
\eea
where $\lfloor \cdot \rfloor$ represents the floor function. 
Similarly, for a rAP node $V_{\mathrm{rAP}, j} \in \mathcal{V}_{\mathrm{rAP}}$, where $j = 1, \ldots, N_{\mathrm{R}}M$, the initial value $\mathbf{g}^{(1)}[V_{\mathrm{rAP}, j}] \in \mathbb{C}^{N_{\mathrm{T}}M}$ is characterized by the edge features associated with the $j$-th rAP node, which can be written as
\bea
&&\hspace{-0.6cm}\mathbf{g}^{(1)}[V_{\mathrm{rAP}, j}] = \Big[\widetilde{\mathbf{A}}_{1, \lfloor \frac{j}{M} \rfloor} (1, j - M\lfloor (j-1)/M \rfloor), \cdots, \nonumber \\
&& \hspace{-0.2 cm}\widetilde{\mathbf{A}}_{1,\lfloor \frac{j}{M} \rfloor} (M, j - M\lfloor (j-1)M \rfloor), ~\widetilde{\mathbf{A}}_{2,\lfloor \frac{j}{M} \rfloor} (1, j - M\lfloor (j-1)/M \rfloor), \nonumber \\
&& \hspace{0.6 cm} \cdots,  \widetilde{\mathbf{A}}_{N_{\mathrm{T}}, \lfloor \frac{j}{M} \rfloor} (M, j- M\lfloor (j-1)/M \rfloor)\Big]^T.
\eea
For a UE node $V_{\mathrm{UE}, k} \in \mathcal{V}_{\mathrm{UE}}$, where $k = 1, \ldots, K$, the initial value $\mathbf{g}^{(1)}[V_{\mathrm{UE}, k}] \in \mathbb{C}^{N_{\mathrm{T}}M}$ is given by
\bea
\mathbf{g}^{(1)}[V_{\mathrm{UE}, k}] = \left[\mathbf{h}_{1, k}(1), \cdots, \mathbf{h}_{1, k}(M), \cdots, \mathbf{h}_{N_{\mathrm{T}}, k}(M)\right]^T.
\eea

After initialization, for a node $V_t \in \mathcal{V}$ with its hidden state vector being $\mathbf{g}^{(l)}[V_t]$ in the $l$-th ($l>1$) layer, our goal is to extract the important features from its neighbors $V_s \in \mathcal{N}(V_t)$ and update its hidden state for the next layer, i.e., $\mathbf{g}^{(l+1)}[V_t]$, through a transformer-based heterogeneous message passing scheme.  By applying a residual connection, the updated hidden state $\mathbf{g}^{(l+1)}[V_t]$ can be expressed as follows:
\vspace{-0cm}
\be
\mathbf{g}^{(l+1)}[V_t] = f_{\mathrm{agg}}\Big(f_{\mathrm{att}}\big(V_t,\mathbf{g}^{(l)}[V_t]\big) \cdot \underset{V_s\in\mathcal{N}(V_t)}{f_{\mathrm{msg}}}\big(\mathbf{g}^{(l)}[V_s]\big)\Big) + \mathbf{g}^{(l)}[V_t], \vspace{-0 cm}
\ee
where $f_{\mathrm{agg}}(\cdot)$, $f_{\mathrm{att}}(\cdot)$, and $f_{\mathrm{msg}}(\cdot)$ represent the functions for aggregation, attention calculation, and message calculation, respectively. The function $f_{\mathrm{msg}}(\cdot)$ is responsible for extracting the message carried by the neighbors of node $V_t$. The importance of each neighboring node is evaluated through function $f_{\mathrm{att}}(\cdot)$. Finally, the weighted message of the set of neighbors is aggregated through $f_{\mathrm{agg}}(\cdot)$
to obtain the updated hidden state for node $V_t$. 

We use a multi-head self-attention mechanism in function $f_{\mathrm{att}}(\cdot)$. Considering $N_{\mathrm{h}}$ self-attention heads, for the $h$-th attention head, we map the hidden state $\mathbf{g}^{(l)}[V_t]$ of node $V_t \in \mathcal{V}$ into the $h$-th query vector $\mathbf{q}^{(l)}_h[V_t]$ through a linear projector $\mu_{\mathrm{q}}(\cdot~; \mathbf{W}_h^{\mathrm{q}})$ with a weight matrix $\mathbf{W}_h^{\mathrm{q}}$. Similarly, the hidden state $\mathbf{g}^{(l)}[V_s]$ of a neighboring node $V_s \in\mathcal{N}(V_t)$ with type $\psi(V_s)$ is projected into the $h$-th key vector $\mathbf{k}^{(l)}_h[V_s]$ with a linear projector $\mu_{\mathrm{k}, \psi(V_s)}(\cdot~; \mathbf{W}^{\mathrm{k}}_{h, \psi(V_s)})$. Note that $\mathbf{W}^{\mathrm{k}}_{h, \psi(V_s)}$ is the weight matrix and different types of nodes have their own unique projectors with different weight matrices. Then, we calculate the similarity between the query and the key as the product $\big(\mathbf{k}^{(l)}_h[V_s]\big)^T \mathbf{W}_{\psi(V_s)}^{\mathrm{att}}\mathbf{q}^{(l)}_h[V_t]$, where $\mathbf{W}_{\psi(V_s)}^{\mathrm{att}}$ is the weight matrix. 
By concatenating all the $N_{\mathrm{h}}$ self-attention heads together, we can obtain the overall attention vector for node $V_t$. 
The function $f_{\mathrm{att}}(\cdot)$ can be expressed as follows:
\vspace{-0 cm}
\bea
&& \hspace{-1.1 cm} f_{\mathrm{att}}\big(V_t,\mathbf{g}^{(l)}[V_t]\big) =  \underset{V_s \in \mathcal{N}(V_t)}{\mathrm{softmax}}\Big(\underset{h}{\big\|}\big\{\big(\mathbf{k}^{(l)}_h[V_s]\big)^T\mathbf{W}_{\psi(V_s)}^{\mathrm{att}}\mathbf{q}^{(l)}_h[V_t]\big\}\Big), \label{eq:att}\\
&& \hspace{0.cm} \mathbf{q}^{(l)}_h[V_t] = \mu_{\mathrm{q}}(\mathbf{g}^{(l)}[V_t]; \mathbf{W}_h^{\mathrm{q}}), \\
&& \hspace{0. cm} \mathbf{k}^{(l)}_h[V_s] = \mu_{\mathrm{k}, \psi(V_s)}\big(\mathbf{g}^{(l)}[V_s]; \mathbf{W}^{\mathrm{k}}_{h, \psi(V_s)}\big), \vspace{-0cm}
\eea 
where $\underset{h}{\mathbin\Vert}\{\cdot\}$ represents the concatenation operation.

\begin{figure}[t]
\centering
\includegraphics[width=3.4 in]{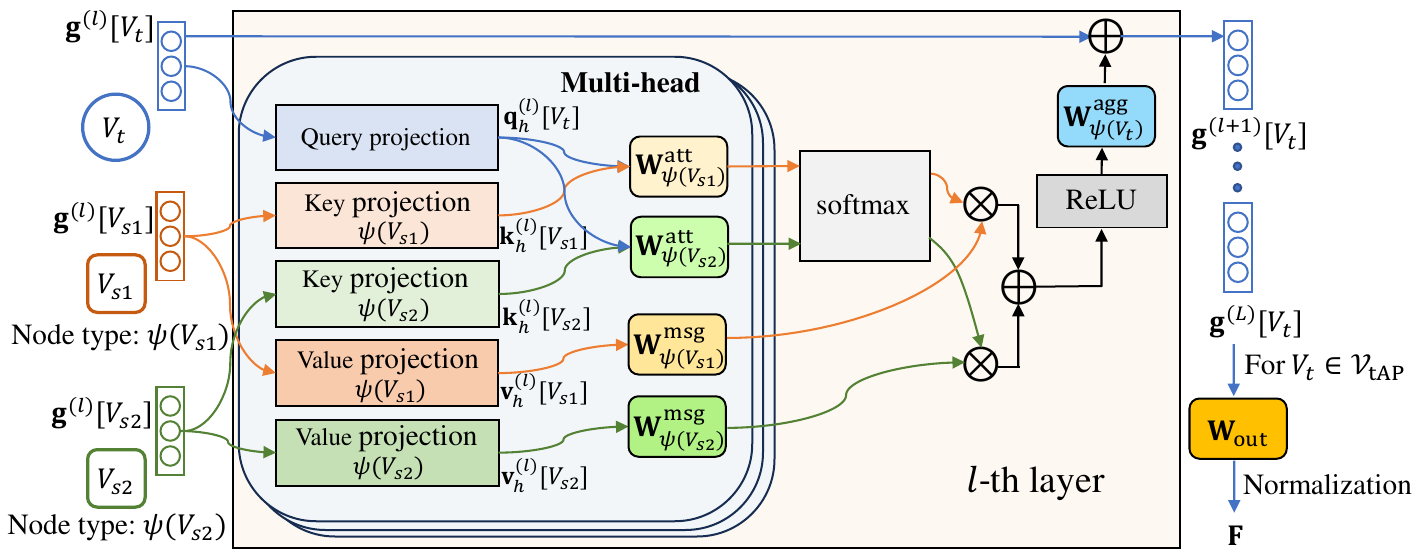}
\vspace{-0.25 cm}
\caption{Transformer-based heterogeneous message passing. Given node $V_t$ and its hidden state $\mathbf{g}^{(l)}[V_t]$, as well as the neighboring nodes $V_{s1}$, $V_{s2} \in \mathcal{N}(V_t)$ and their hidden states $\mathbf{g}^{(l)}[V_{s1}]$ and $\mathbf{g}^{(l)}[V_{s2}]$, the hidden state $\mathbf{g}^{(l+1)}[V_t]$ for node $V_t$ is updated through a transformer-based heterogeneous message passing scheme.}
\Description{Architecture.}
\label{fig:HGT}\vspace{-0.2 cm}
\end{figure}

The information from the neighboring nodes is extracted through function $f_{\mathrm{msg}}(\cdot)$. Similar to $f_{\mathrm{att}}(\cdot)$, for the $h$-th attention head, the hidden state $\mathbf{g}^{(l)}[V_s]$
of a neighboring node $V_s\in \mathcal{N}(V_t)$ is mapped into a value vector $\mathbf{v}^{(l)}_h[V_s]$ through a linear projector $\mu_{\mathrm{v}, \psi(V_s)}(\cdot~; \mathbf{W}_{h, \psi(V_s)}^{\mathrm{v}})$. The multi-head message is calculated by:
\bea
&&\hspace{-1.4 cm}  f_{\mathrm{msg}}\big(\mathbf{g}^{(l)}[V_s]\big) = \underset{h}{\big\|} \big\{ \mathbf{v}^{(l)}_h[V_s]\mathbf{W}_{\psi(V_s)}^{\mathrm{msg}}\big\} \nonumber \\
 &&\hspace{0.55 cm} = \underset{h}{\big\|}\Big\{\mu_{\mathrm{v},\psi(V_s)}\big(\mathbf{g}^{(l)}[V_s]; \mathbf{W}_{h, \psi(V_s)}^{\mathrm{v}}\big)\mathbf{W}_{\psi(V_s)}^{\mathrm{msg}}\Big\}, 
\eea 
where $\mathbf{W}_{\psi(V_s)}^{\mathrm{msg}}$ is the weight matrix. 
Finally, the calculated multi-head attention scores and messages from all the neighboring nodes are aggregated followed by an activation operation. We use the rectified linear unit (ReLU) as the activation function. The hidden state of node $V_t$ in layer $(l+1)$ is then updated as follows:
\bea
&&\hspace{-1 cm} \mathbf{g}^{(l+1)}[V_t] = \mathbf{W}_{\psi(V_t)}^{\mathrm{agg}}\mathrm{ReLU}\Big(\mathrm{sum}\big(f_{\mathrm{att}}(V_t, \mathbf{g}^{(l)}[V_t]) \nonumber \\ && \hspace{1.6 cm} \cdot ~\underset{V_s\in\mathcal{N}(V_t)}{f_{\mathrm{msg}}}(\mathbf{g}^{(l)}[V_s])\big)\Big)  + ~\mathbf{g}^{(l)}[V_t], 
\eea
where  $\mathbf{W}_{\psi(V_t)}^{\mathrm{agg}}$ denotes the weight matrix to update the hidden state of the node $V_t$. We show the heterogeneous message passing scheme in Fig. \ref{fig:HGT}. 
By stacking all the $L$ hidden layers, the constructed SACGNN can generate a representation $\mathbf{g}^{(L)}[V_t]$ for each node $V_t \in \mathcal{V}$. 
The hidden states of the final layer for the tAP nodes, i.e., $\mathbf{g}^{(L)}[V_{t}]$ for $V_{t}\in \mathcal{V}_{\mathrm{tAP}}$, are then used for the transmit beamforming design. We apply a linear transformation of weight $\mathbf{W}_{\mathrm{out}}$  to the hidden states $\mathbf{g}^{(L)}[V_{t}]$ with $V_{t}\in \mathcal{V}_{\mathrm{tAP}}$.  The outputs are normalized to satisfy the power constraint (\ref{eq:power constraint}), and we obtain the designed beamforming matrix for all the transmit APs, i.e.,  $\mathbf{F} \in \mathbb{C}^{N_{\mathrm{T}}M \times (K+1)}$, where the beamforming matrix for the $i$-th transmit AP is given by $\mathbf{F}_i = \mathbf{F}((i-1)M +1: iM)\in \mathbb{C}^{M\times (K+1)}$, $i = 1, \ldots, N_{\mathrm{T}}$. 

\subsection{Loss Function}

We construct a training data sample as $(\mathbf{H}, \widetilde{\mathbf{A}})$, where $\mathbf{H}=\{\mathbf{h}_{i,k}\}_{i,k}^{N_{\mathrm{T}}, K}$ and $\widetilde{\mathbf{A}} = \{\widetilde{\mathbf{A}}_{i,j}\}_{i,j}^{N_{\mathrm{T}}, N_{\mathrm{R}}}$. The training dataset is given by 
$\mathcal{D} = \{(\mathbf{H}^{(1)}, \linebreak \widetilde{\mathbf{A}}^{(1)}), \ldots, (\mathbf{H}^{(N_{\mathrm{tr}})}, \widetilde{\mathbf{A}}^{(N_{\mathrm{tr}})})\}$, where $N_{\mathrm{tr}}$ is the total number of training samples. The developed SACGNN is trained in an unsupervised manner. In particular, we reformulate the objective function in (\ref{eq:original problem}) and define the loss function as follows: 
\bea
\mathcal{L} = - R + \rho\mathrm{ReLU}(\gamma_{\min} - \gamma^{(\mathrm{s})}), \label{eq:loss}
\eea 
where the sensing SNR requirement (\ref{eq:sense constraint}) is moved to the objective with a penalizing coefficient  $\rho$. The developed SACGNN is trained to minimize the loss function (\ref{eq:loss}) using Adam optimizer \cite{Adam} in an unsupervised manner. After training, the trained model can be applied for online beamforming design. When given the communication channel vectors\footnote{We note that channel estimation is an important topic and there exist various estimation methods that can be applied to obtain the channel vectors. For simplicity, we assume perfect channel state information is available at the CPU which is widely adopted in the existing works (e.g., \cite{UD-Asilomar,SL-WCNC,ZB-TWC}).} and AoDs/AoAs of the target to be detected, the network can generate the beamforming vectors for the transmit APs to achieve a high sum-rate while guaranteeing the sensing SNR requirement. 

\section{Performance Evaluation \label{sec:simulation}}

In this section, we evaluate the performance of the developed SACGNN framework for cooperative ISAC beamforming. The simulation settings are as follows. We consider there are $N_{\mathrm{T}} = 2$ transmit APs and $N_{\mathrm{R}} =2$ receive APs to cover a $200 \times 200$ m$^2$ area. Considering a three-dimensional (3D) $[\mathrm{x}, \mathrm{y}, \mathrm{z}]$ coordinate, the transmit APs are located at $[0,100,20]$ and $[200,100,20]$. The receive APs are placed at $[100,0,20]$ and $[100,200,20]$. Each AP is equipped with a UPA of $M_{\mathrm v}=4$ vertical antennas and  $M_{\mathrm{h}}=16$ horizontal antennas. We consider there are $K=4$ users. The users and the target are assumed to be randomly distributed in the area,
with $\mathrm{x}$ and $\mathrm{y}$  coordinates ranging from $0$ to $200$, and the $\mathrm{z}$  coordinate ranging from $0$ to $35$.
We adopt the normalized system parameters \cite{UD-Asilomar}. In particular, the variance of communication channel gain $\zeta^2_{i,k}$ is set to $0.5, i=1, \ldots N_{\mathrm{T}}, k=1, \ldots, K$. The variance of communication noise $\sigma^2_k$ at the $k$-th user, $k=1, \ldots, K$, and the variance of sensing noise $\xi^2_j$ at the $j$-th receive AP, $j=1, \ldots, N_{\mathrm{R}}$, are set to $1$. The variance of the sensing channel gain 
$\chi^2_{i,j}$ is set to $0.1, i=1, \ldots, N_{\mathrm{T}}, j=1, \ldots, N_{\mathrm{R}}$. We set the number of hidden layers $L$ to two. The learning rate is set to $10^{-4}$. 
We generate $10,000$ data samples, where $8,000$ of them are used for offline training and $2,000$ are used for testing. 
We compare the performance of our proposed SACGNN framework with two other baseline methods: (i) null-space projection based beamformer for sensing and regularized zero-forcing beamformer for communication  (NS-RZF) proposed in \cite{UD-Asilomar}; (ii) a DNN-based scheme which consists of three 3D convolutional neural network layers followed by a fully-connected layer.

\begin{figure}[t]
\centering
\includegraphics[width=2.2 in]{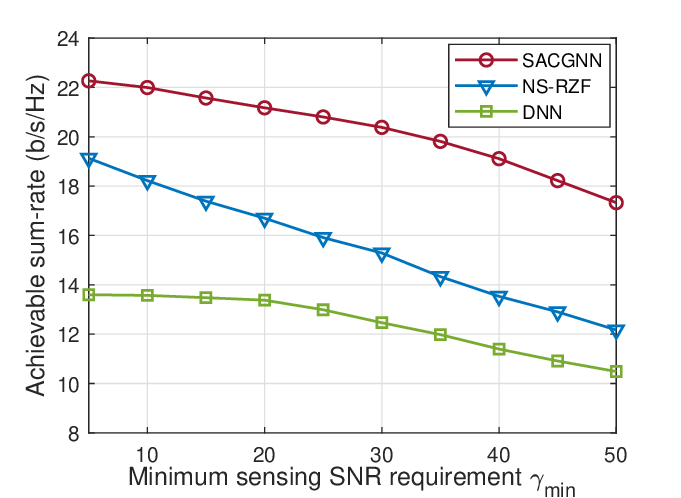}
\vspace{0mm}
\caption{The achievable sum-rate versus the minimum sensing SNR requirement $\gamma_{\min}$, where $P_{\max}=30$ dBm.}\vspace{-0.3 cm}
\label{fig:rate vs SNR}
\end{figure} 

In Fig. \ref{fig:rate vs SNR}, we show the achievable sum-rate under different sensing SNR requirements. The maximum transmit power $P_{\max}$ is set to $30$ dBm. We can observe that as the sensing requirement becomes more stringent, the achievable  sum-rate decreases since more power is allocated to the sensing beamformer. The results demonstrate that the proposed SACGNN framework outperforms the NS-RZF scheme proposed in \cite{UD-Asilomar}. The NS-RZF scheme designs the sensing beamformer to be in the null space of the communication channels and manages inter-user interference through regularized zero-forcing. This scheme may not perform well since the communication channels of users and the target are usually correlated with each other, leading to reduced null space of the communication channels. 
Moreover, the proposed SACGNN framework also outperforms the DNN-based baseline approach. This is due to the fact that the beamforming design needs to simultaneously address both communication and sensing features, leading to additional complexity in the design process. Typical 3D convolutional neural networks may not be sufficient to capture the intricate interplay between communication and sensing information. The results in Fig. \ref{fig:rate vs SNR} demonstrate the performance gain of the proposed SACGNN framework. 

\begin{figure}[t]
\centering
\includegraphics[width=2.4 in]{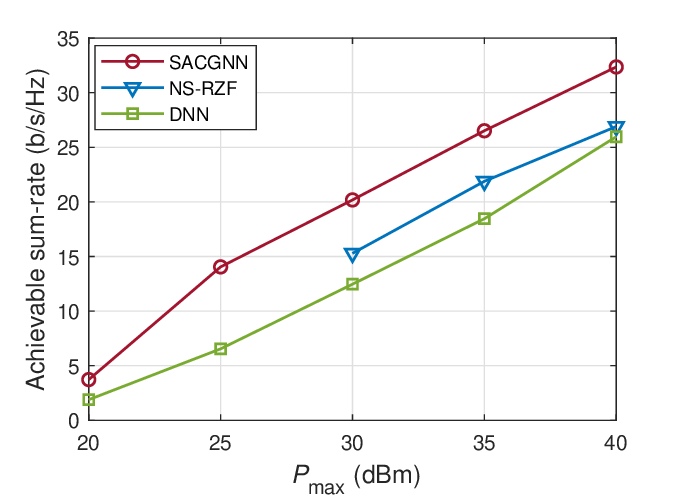}
\vspace{0mm}
\caption{The achievable sum-rate versus the maximum transmit power $P_{\max}$, where $\gamma_{\min} = 30$.}\vspace{-0.3 cm}
\label{fig:rate vs P}
\end{figure} 

In Fig. \ref{fig:rate vs P}, we show the achievable sum-rate versus the maximum transmit power, under a fixed sensing SNR requirement, i.e., $\gamma_{\min}=30$.  Similarly, we observe that the proposed approach consistently provides a higher sum-rate compared to the other two baseline methods. In particular, in the low transmit power regime ($P_{\max} = 20, 25$ dBm), the NS-RZF scheme fails to satisfy the sensing SNR requirement. This is because the sensing and communication channels exhibit high correlations, resulting in a limited null space dimension of the communication channels and consequently a marginal sensing beamforming gain.

\begin{figure}[t]
\centering
\subfigure[A 3D view of the beam pattern.]{
\includegraphics[height=1.43in]{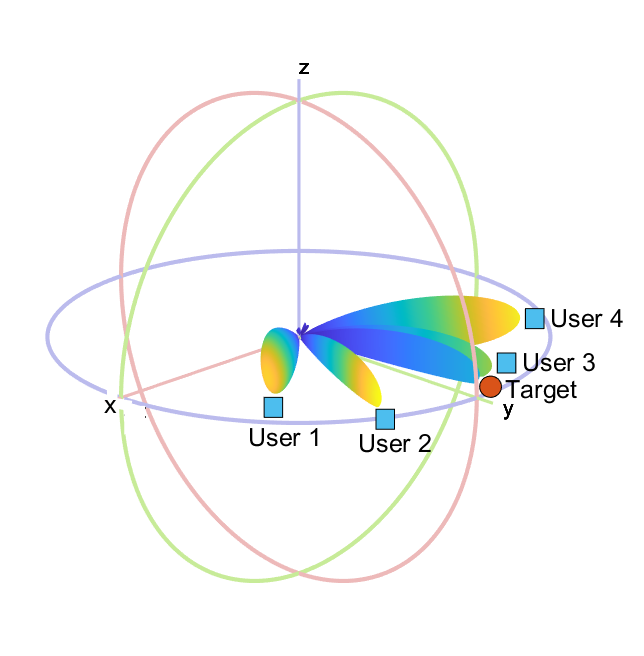}} \hspace{-0.3 cm}
\subfigure[The azimuth cut of the beam pattern.]{
\includegraphics[height=1.43in]{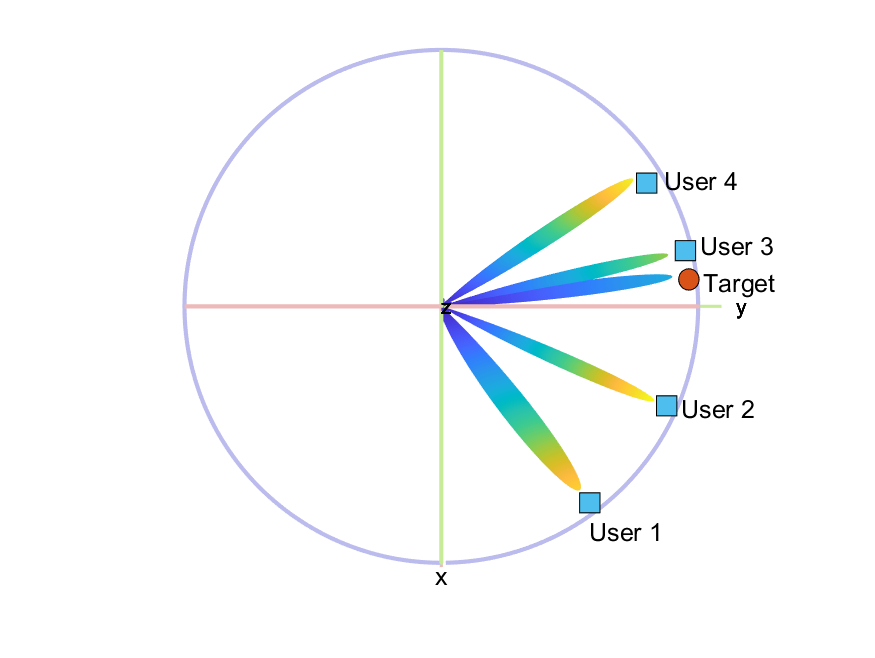} }

\caption{Beam patterns captured at a transmit AP located at $[0, 100,20]$. } 
\label{fig: beam pattern}
\vspace{-0.2 cm}
\end{figure}

Finally, the beam patterns generated by one of the transmit APs using our proposed SACGNN framework are shown in  Fig. \ref{fig: beam pattern}, where a 3D view and the azimuth cut are presented. The results are generated based on the following settings. The four users are located at $[40, 40, 30]$, $[140, 40, 20]$, $[140, 140, 20]$,  and $[40, 140, 30]$, respectively. The target is located at $[115, 115, 25]$. The view is captured from a transmit AP located at $[0, 100, 20]$.
It can be observed from Fig. \ref{fig: beam pattern} that the generated beamformer can effectively manage the side lobes and interference, even when the target to be detected is close to a communication user.

\section{Conclusion \label{sec:conclusion}}

In this paper, we investigated the cooperative ISAC beamforming design in cell-free MIMO systems. We proposed a SACGNN framework which models the cell-free MIMO system for cooperative ISAC as a heterogeneous graph and uses a transformer-based scheme for heterogeneous message passing. We conducted simulations for performance evaluation and included two baseline methods for comparison. Simulation results showed that the proposed SACGNN framework outperforms a conventional null-space projection based scheme and a DNN-based baseline scheme. Moreover, results 
demonstrated that the proposed SACGNN framework can generate narrow beamformers toward users and the target, where the side lobes can be effectively managed.

\begin{acks}
The work is supported in part by the Government of Canada Innovation for Defence Excellence and Security (IDEaS) program and the Digital Research Alliance of Canada (alliancecan.ca).
\end{acks}


\bibliographystyle{ACM-Reference-Format}
\bibliography{ISAComref}

\end{document}